\documentclass{iopjournal}

\usepackage{graphicx} 
\usepackage[style=phys]{biblatex}

\usepackage{amsmath,amsfonts,amssymb}
\usepackage{mathrsfs}
\usepackage{float}
\usepackage{physics}

\usepackage[normalem]{ulem}		
\usepackage{color}
 
\newcommand{\kB}{k_{\mathrm{B}}}

\newcommand{\e}{\mathrm{e}}
\newcommand{\myfrac}[3][0pt]{\genfrac{}{}{}{}{\raisebox{#1 pt}{$#2$}}{\raisebox{-#1 pt}{$#3$}}}

\begin{document}

\title{Mpemba effect without a wall}

\author{Siddharth S. Sane$^1$ and John Bechhoefer$^1$}

\affil{$^1$Department of Physics, Simon Fraser University, Burnaby BC V5A 1S6, Canada}

\email{siddharth\_sane@sfu.ca, johnb@sfu.ca} 


\begin{abstract}
Experimental demonstrations of the Mpemba effect in a colloidal-particle system have imposed an instantaneous temperature quench via an initial condition where individual particle trajectories are drawn from a high-temperature  Boltzmann distribution and then evolve in a bath at lower temperature.  The potential used for the high-temperature distribution has had two walls that impose a finite range of initial positions, even at effectively infinite temperatures.  The potential for evolution in the bath had no walls and matched that used for the initial condition over its support.  For low bath temperatures, the difference between the dynamics with and without walls is negligible.  Nonetheless, it has been speculated that the walls are perhaps more important than they seem and might even be required for the Mpemba effect to occur.  Here, we introduce a new potential that lacks walls and use it for both the initial high temperature state and for subsequent evolution. This new potential then allows us to investigate by simulation Mpemba effects at finite-quench rates.  Our results help clarify intuitions concerning the existence of a Mpemba effect and show that the Mpemba effect appears only for bath-temperature quenches faster than a critical quench rate.
\end{abstract}

\section{Introduction}

Although qualitative discussions go back as far as Aristotle in Classical Greece~\cite{Webster1923aristotle}, the Mpemba effect in its modern form was first described in a paper by Erasto Mpemba and Denis Osborne in 1969~\cite{Mpemba1969}.  They observed that water at boiling temperatures froze more rapidly than the same volume of room-temperature water. The effect defies common sense, as hot water would seemingly have to first cool to room temperature before it could freeze.  Although the existence of the Mpemba effect in freezing water remains controversial~\cite{Burridge2016, Burridge2020,Brownridge2025a}, it has been observed in a variety of other settings, including clathrate hydrates~\cite{Ahn2016}, metallic glasses~\cite{Song2026}, magnetoresistant magnetites~\cite{Chaddah2010}, polylactide crystallisation~\cite{Hu2018}, and simulations of granular fluids~\cite{Lasanta2017}.  The ubiquity of the Mpemba effect across such a diverse set of systems motivates a search for a generic explanation that goes beyond system-specific scenarios.

In 2017, Lu and Raz  proposed the first such explanation of the Mpemba effect~\cite{Lu2017}.  Their theory assumed Markovian dynamics and considered a system that was first in equilibrium at a temperature $T_0$ and then suddenly quenched in a bath at constant temperature $T_b$.  For overdamped dynamics of a continuous state $x$ described by Langevin equations, they focused evolution of the probability density function $p(x,t)$.  At long times, the approach to the equilibrium Boltzmann distribution $\pi(x;T_b)$, described by the Fokker-Planck equation~\cite{Risken1996} is a single exponential,
\begin{align}
	p(x,t) \sim \pi(x;T_b) + a_2 v_2(x;T_b) \, \e^{-\lambda_2 t} \,,
\label{eq:density_asymptote}
\end{align}
where $\lambda_2 = \lambda_2(T_b) > 0$ is the slowest relaxation rate among an infinite spectrum of eigenvalues $\lambda_2 < \lambda_3 < \cdots$, the corresponding right eigenfunction for dynamics evolving at the bath temperature is $v_2(x;T_b)$, and the coefficient $a_2$ is proportional to the projection of the initial equilibrium state (Boltzmann distribution at temperature $T_0$) on the system dynamics at the bath temperature $T_b$, as represented by the left eigenfunction $u_2(x;T_b)$:
\begin{align}
	a_2(T_0,T_b) \propto \int_{-\infty}^\infty \dd{x} \pi(x;T_0) \, u_2(x;T_b) \,.
\label{eq:a2}
\end{align}

The projection coefficient $a_2$ must equal zero when the initial and final temperatures match, $T_0=T_b$, since the system is always in equilibrium.  The intuition from slow cooling described above predicts that $|a_2|$ should monotonically increase with the initial temperature $T_0 > T_b$.\footnote{
Because the sign of $a_2$ can be positive or negative, it is the dependence of the absolute value of $a_2$ on the initial temperature $T_0$ that determines whether cooling is normal or anomalous.} 
However, Lu and Raz observed that $|a_2(T_0,T_b)|$ need not be monotonic and associated its decrease, $\dv*{|a_2|}{T_0} < 0$, with the appearance of a (regular) Mpemba effect.

\begin{figure}
 \centering
        \includegraphics[width=0.9\textwidth]{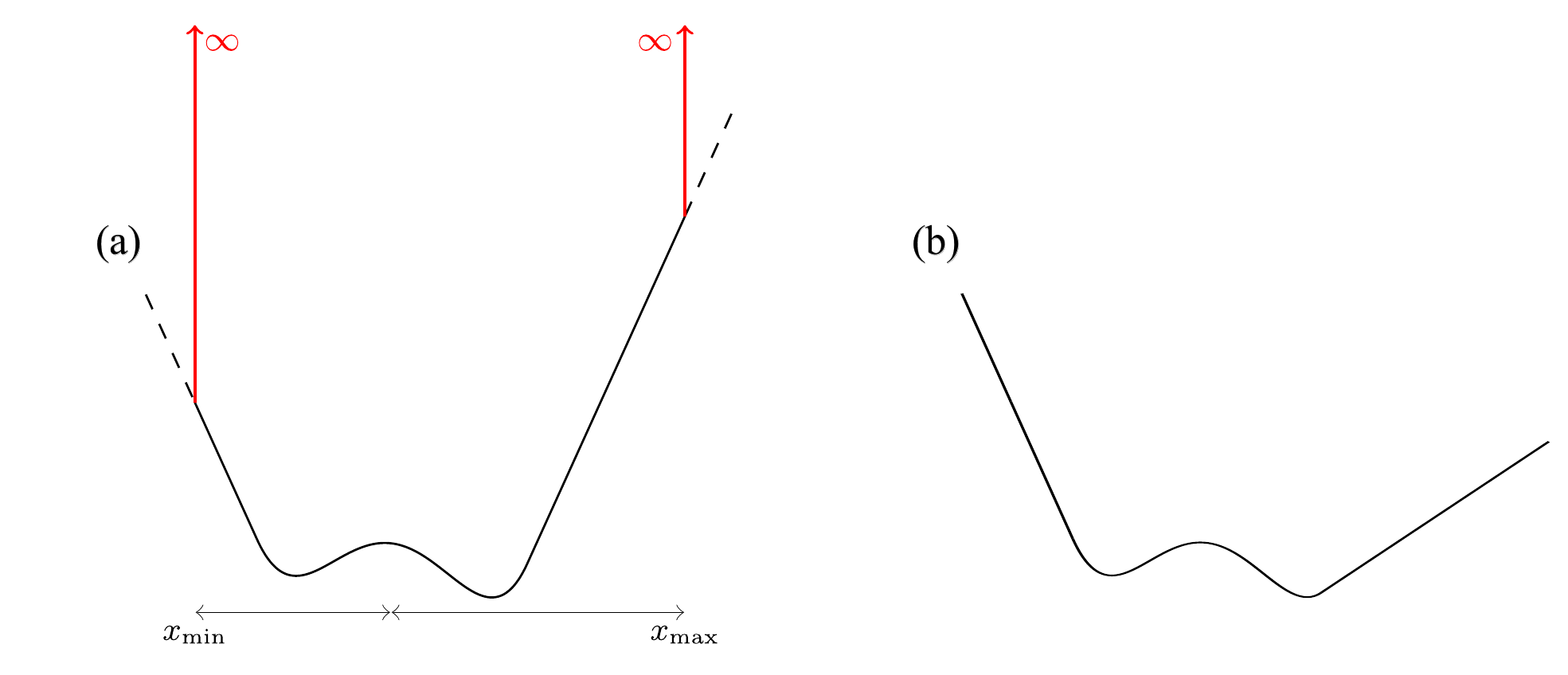}
 \caption{Potentials showing a Mpemba effect.  (a) Kumar-Bechhoefer potentials~\cite{Kumar2020}.  The potential used for the Boltzmann distribution at $T_0$ has vertical walls (red) at $x_\mathrm{min}$ and $x_\mathrm{max}$, whereas the potential used for the evolution in the bath has no walls and extends beyond (dashed lines).  (b)  The potential used here has no walls.  The different colors used in the potentials denote regions where the function is analytic; at color changes, either the first or second derivative is discontinuous.  The central portion of the potentials $U_0(x)$ is a tilted double-well potential with energy barrier $E_b$ and energy difference $\Delta E$ between wells. 
}
\label{fig:potOldNew}
\end{figure}

In 2020, Kumar and Bechhoefer proposed the first experimental test of the Lu-Raz theory, in the context of single-particle colloid dynamics in water, using a carefully designed external potential~\cite{Kumar2020,Kumar2021},  figure~\ref{fig:potOldNew}(a).  This paper demonstrated not only the basic scenario outlined above but also found a special case where $a_2(T_0,T_b) = 0$ for a specific temperature $T_0>T_b$, a situation known as the \textit{strong} Mpemba effect~\cite{Klich2019}.  In that case, the vanishing $a_2$ coefficient meant that the long-time dynamics decayed at a rate $\lambda_3 > \lambda_2$ and thus was exponentially faster than the generic case, where $a_2 \neq 0$.  Compared to the case of complicated systems such as water-and-ice, the colloidal dynamics was simple to analyze quantitatively and featured much faster dynamics ($\approx 0.1$~s~/~trial, compared to $\approx 1$~hr/~trial in water).  Experiments with $10^3$ to $10^4$ trials were then straightforward, which was at least 100 times greater than the number of trials in the experiments on water.

The success of this theory and initial experiments led to many further investigations of Mpemba-like effects, including related work on suspended nanoparticles in a vacuum chamber~\cite{Tian2025}, and to various attempts to generalize the effect, such as inverse Mpemba effects (anomalous heating)~\cite{Lu2017,Lasanta2017,Kumar2022}, multiple transitions~\cite{Klich2019,malhotra2024double}, asymmetries between heating and cooling~\cite{lapolla2020faster,ibanez2024heating,Dieball2026}, dynamical phase transitions while heating and cooling~\cite{meibohm2024exponential,Meibohm2026}, transitions to and/or from nonequilibrium steady states~\cite{degunther2022anomalous,teza2023relaxation}.  Perhaps the most notable generalizations have been to the quantum domain, an effort that has generated considerable activity~\cite{ares2025quantum}.  For a recent comprehensive review of anomalous relaxation, including Mpemba and related effects, see reference~\cite{Teza2026}.

The experimental studies on colloidal systems suffered from a flaw that traces back to experimental limitations.  As discussed above, the instantaneous quench is imposed by first drawing initial positions from a Boltzmann distribution in equilibrium at $T_0$ and then evolving the system in a bath at $T_b$.  In reference~\cite{Kumar2020}, the potentials in these two steps were actually slightly different:  As illustrated in figure~\ref{fig:potOldNew}(a), the initial distribution was drawn from a tilted double potential truncated by vertical walls at positions $-x_\mathrm{min}$ and $x_\mathrm{max}$, whereas the physical potential has no walls. (We define $x_\mathrm{min}>0$ for convenience.)  The bath potential coincides with the initial-value potential for $x \in [- x_\mathrm{min}, x_\mathrm{max}]$ but then extends beyond (dashed lines).  The walls were necessary for the initial-value potential because of the need to confine the initial density $\pi(x,T_0)$ within the maximum experimental domains, even at extremely high initial temperatures.  Because the bath temperature $T_b \ll T_0$, any particle that is near $-x_\mathrm{min}$ or $x_\mathrm{max}$ quickly moves in towards the closest potential well and has very little probability to fluctuate outside the initial domain. As a consequence, the effect of using two different potentials for the two regimes was negligible.

Although the practical consequences of using two different potentials in reference~\cite{Kumar2020} were small in practice, one might worry whether the walls play a more important role than implied. Indeed, two recent (related) papers argue that a box can play a key role in enabling a Mpemba effect and might even be a ``necessary condition'' for the effect~\cite{Liu2026a, Liu2026}.  Here, we will show through an explicit counterexample, depicted in figure~\ref{fig:potOldNew}(b), that this is not the case.

A second motivation for finding potentials with no walls that show a Mpemba effect is a desire to generalize the cooling protocol for the relaxation from one with an infinite quench rate to one with a finite quench rate.  Such a generalization is interesting for two reasons:

\begin{enumerate}
\item  Conceptually, we know that the Mpemba effect cannot exist in a system undergoing a slow (quasistatic) quench.  In that case, the system passes through all intermediate temperatures, implying that the equilibration time must increase monotonically with the initial system temperature.  At the same time, we have specific cases where Mpemba effects have been demonstrated for protocols with infinite quench rates (by experiment, simulation, and analytic work).  Since these same systems cannot show a Mpemba effect when quenched very slowly, it follows that there should be a critical quench rate required to observe a Mpemba effect.  How fast is this rate?

\item Practically, an effectively infinite quench rate can only be imposed in very special cases, such as the colloidal single-particle system described above.  In extended materials, as discussed later, there will be a characteristic time to change temperature that scales with system size.  Thus, even if a Mpemba effect is in principle possible, it may not be possible to quench a particular system fast enough.  Again, knowing the quench rate required for the Mpemba effect would influence designs needed to produce it.
\end{enumerate}

Implementing a finite quench experimentally requires changing the way the way effective temperatures are imposed.  The initial-value method discussed above is suitable only for infinite-rate quenches.  To impose time-dependent quenches, we will add artificial noise~\cite{Martinez2013,Chupeau2018}.  An advantage of such a technique is that the potential used for the initial condition is automatically the same as the one used for the subsequent quench, resolving the ``two-potential" issue.  A problem, though, is that the high-temperature Boltzmann distribution must have (almost) all of its probability within the experimentally accessible domain.  In reference~\cite{Kumar2020}, the initial temperature was as high as $T_0 = 1000~T_b$, which would imply a domain size that far exceeds what is possible experimentally.  Thus, a new form of potential requiring lower $T_0$ values is needed to explore finite dynamics.  Indeed, this was our original motivation for searching for a ``potential without walls" for the Mpemba effect.

In this paper, we will

\begin{enumerate}
\item Present a one-dimensional potential for single, overdamped particle dynamics that is defined for $x \in (-\infty,+\infty$) (has no walls) and yet shows a Mpemba effect for $T_0 \approx 10~T_b$.  The new potential, depicted in figure~\ref{fig:potOldNew}(b), allows us to clarify intuitions about the Mpemba effect.

\item  Resolve apparent contradictions between our result and the conclusions of references~\cite{Liu2026a, Liu2026}, which would seem to rule out such a potential.

\item  Carry out preliminary simulations showing that the newly designed potential can display a Mpemba effect and that it is possible to observe the associated critical (finite) quench rate. 
\end{enumerate}

In section~\ref{sec:background}, we review the Mpemba theory and our experimental results in more detail, to motivate more precisely the need for a new potential. In section~\ref{sec:walls}, we discuss how walls can lead to Mpemba effects.  In section~\ref{sec:newPotential}, we present a  potential without walls that leads to a Mpemba effect.  In section~\ref{sec:experimental}, we discuss the challenge of finding parameters for the potential that are compatible with experimental limitations.  In section~\ref{sec:analytic}, we discuss the role of the analytic structure of the potential.  This helps explain the apparent necessity of walls and also justifies the use of functions with discontinuous second derivatives in experimental studies.  Then, in section~\ref{sec:finiteRate}, we present preliminary simulations and experimental results for finite-rate quenches.  

This paper is largely drawn from the MSc~thesis of Siddharth Sane~\cite{Sane2026}.

\section{Background}
\label{sec:background}

We focus on one-dimensional overdamped dynamics where trajectories $x(t)$ are described by a Langevin equation 
\begin{align}
	\dot{x}(t) =\frac{1}{\gamma} F(x(t)) + \sqrt{2 D} \, \eta(t) \,,
	\label{eq:langevin}
\end{align}
with diffusivity $D = \kB T_b / \gamma$ constant for constant bath temperature $T_b$; here, $\kB$ is Boltzmann's constant and $\gamma$ is the local friction coefficient.  In one spatial dimension, the force field $F(x)$ can always be written as the negative gradient of a potential $U(x)$, so that $F(x) = -\partial_x U(x)$.  The stochastic force $\eta(t)$ results from the random thermal kicks felt by a Brownian particle in a fluid and is modeled as Gaussian white noise, with $\langle \eta(t) \rangle = 0$ and $\langle \eta(t) \, \eta(t') \rangle = \delta(t-t')$.

The probability density of states $p(x,t)$ at a time $t$ after the initiation of the temperature quench obeys a Fokker-Planck equation (FPE),
\begin{align}
	\pdv{p}{t} = -\frac{1}{\gamma}\pdv{x}[F(x) \, p(x,t)]+D\pdv[2]{p}{x} \,,
\label{eq:FPE}
\end{align}
with initial and final conditions that are Boltzmann distributions $ \pi(x;T) \propto \e^{-U(x) / \kB T}$ for temperatures $T=T_0$ and $T_b$, respectively:
\begin{subequations}
\begin{align}
    p(x,0) &= \pi(x;T_0) \\
    \lim_{t\rightarrow\infty} p(x,t) &= \pi(x;T_b) \,.
\end{align}
\end{subequations}
These conditions ensure that the system cools from $T_0$ to $T_b$. 

One way to solve the Fokker-Planck equation is through an eigenfunction expansion, which gives
\begin{align}
	p(x,t) = \pi(x;T_b) + \sum_{n=2}^\infty a_n(T_0,T_b) \, v_n(x;T_b) \, \e^{-\lambda_n t} \,,
\label{eq:FPEsoln}
\end{align}
where the eigenvalues $\lambda_n(T_b) > 0$ increase with $n$, the $v_n$ are the corresponding right eigenfunctions, and the coefficients $a_n$ are given by
\begin{align}
	a_n(T_0,T_b)   &= \myfrac[2]{\int_{-\infty}^\infty \dd{x} \pi(x;T_0) \, u_n(x;T_b)}{\int_{-\infty}^\infty \dd{x} u_n(x;T_b) \, v_n(x;T_b)} \,,
\label{eq:an}
\end{align}
with $u_n$ the left eigenfunctions corresponding to $\lambda_n$.  For long times, the expression for $p(x,t)$ in equation~\eqref{eq:FPEsoln} approaches the one given in equation~\eqref{eq:density_asymptote}.

Although $p(x,0)$ and $p(x,t \to \infty)$ are both given by Boltzmann distributions, the probability density at intermediate times after an instantaneous quench is in general not a Boltzmann distribution for any temperature.  Nonetheless, one can use a distance function $\mathcal{D}[p(x,t);\pi(x;T_b)]$ to quantify how ``far'' the probability density $p(x,t)$ is from its long-time, equilibrium form.  Typical choices of distance functions include the Kullback-Leibler divergence, which is connected to free-energy differences~\cite{shaw84,Chetrite2021}, and the $L_1$ norm, which is easy to compute.  We use the latter functional to define the time-dependent function
\begin{align}
	\mathcal{D}_{L_1}[p,\pi](t) = \int_{-\infty}^\infty \dd{x} |p(x,t) - \pi(x;T_b)| \,,
\label{eq:L1}
\end{align}

\begin{figure}
 \centering
        \includegraphics[width=0.9\textwidth]{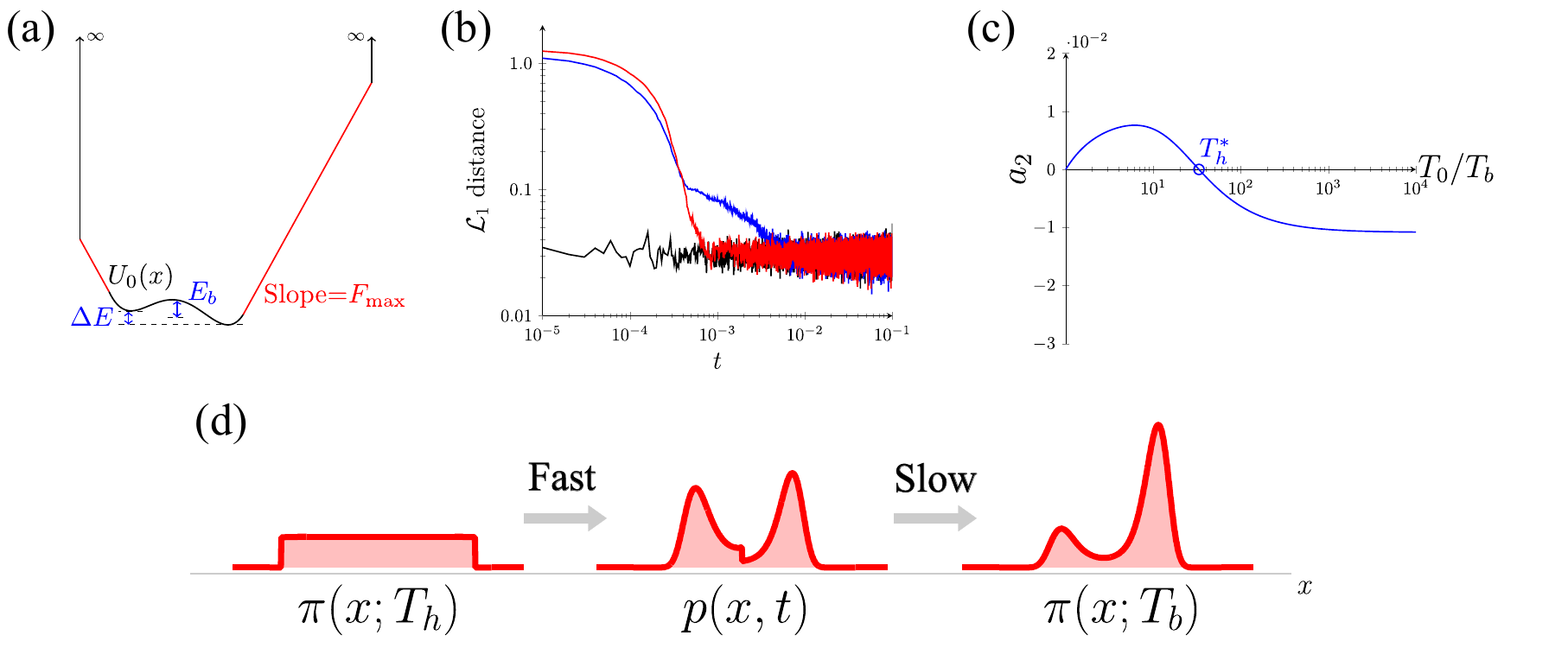}
 \caption{Mpemba scenario from reference~\cite{Kumar2020}. (a) Potential $U(x)$ showing central tilted double-well portion $U_0(x)$ in black, constant force $F_\mathrm{max}$ in red.  Vertical walls are imposed for the initial condition, whereas the particle moves in the potential where $\pm F_\mathrm{max}$ is applied to all $x$.  (b)  Crossing of distance curves indicates the Mpemba effect.  (c)  Projection coefficient $a_2$ as a function of initial temperature $T_0$.  The decrease in $|a_2|$ in $T_0 \in (3,100)$ indicates the Mpemba effect.  The zero crossing at $T_0 = T_h^*$ indicates the strong Mpemba effect.  (d)  Illustration of the two-stage relaxation mechanism.  The initial condition (left) is a Boltzmann distribution for $T_0 = 1000 T_b$ and is nearly a uniform distribution, truncated by two walls.  Individual particles mostly relax quickly to the nearest well (center).  At long times, the distribution approaches the equilibrium Boltzmann distribution for $T_b$. Part (d) is adapted from reference~\cite{,Chetrite2021}.
 } 
\label{fig:KBscenario}
\end{figure}

Using such a distance function, the Mpemba effect can be inferred at finite times from the crossing of distance curves, as shown in figure~\ref{fig:KBscenario}(b).  Within a wide class of distance functions, observing a Mpemba effect using one distance function guarantees that it will be present using other distance functions~\cite{Lu2017,Klich2019,Vu2025}. 

In the colloidal experiments~\cite{Kumar2020}, the low-energy section of the potential was given by a tilted double-well potential, with energy barrier $E_b$ located at $x=0$ (for no tilt) and energy difference $\Delta E$. 

For low bath temperatures, where $\kB T_0 \ll E_b$, the system behaves approximately as a coarse-grained two-state system, with microscopic states confined to the left or right wells.  The equilibrium ratio of probabilities to be in the left state relative to being in the right state is then given by $\e^{-\Delta E / \kB T_b} = \Pi_L(T_b) / \Pi_R(T_b)$, where 
\begin{align}
	\Pi_L = \myfrac[2]{\int_{-\infty}^0 \dd{x} p(x,t)}{\int_{-\infty}^\infty \dd{x} p(x,t)}
\end{align} 
is the probability for a system to be in the left well and $\Pi_R = 1-\Pi_L$ is the probability to be in the right well.  We will use the notation $\Pi_L(T)$ to denote the particular case where $p(x,t) = \pi(x;T)$.  In other words, $\Pi_L(T)$ is the equilibrium probability to be in the left well for temperature $T$.

For high temperatures, where $\kB T_0 \gg E_b$, the relative probabilities for the two states $x \gtrless 0$ are given by the high-energy behavior of the potential $U(x)$.  In reference~\cite{Kumar2020}, these probabilities were asymptotically set by the wall positions  $x = \{ -x_\mathrm{min},x_\mathrm{max} \}$, with $\Pi_L(T_0) = x_\mathrm{max} / L$ and $\Pi_R(T_0) = x_\mathrm{min} / L$, where $L = x_\mathrm{min} + x_\mathrm{max}$ is the total domain size.

The existence of the Mpemba effect was then argued to be a consequence of a two-stage dynamics that started from a high-temperature state and quickly relaxed first to a local equilibrium in which  $\Pi_L / \Pi_R \approx x_\mathrm{min} / x_\mathrm{max}$ is controlled by the well positions and from there to a value $\approx \e^{-\Delta E / \kB T_b}$ controlled by the energy difference $\Delta E$.  The two-stage relaxation is illustrated schematically in figure~\ref{fig:KBscenario}(d).  Generically, the probability ratio $\Pi_L / \Pi_R$ set by the walls is not equal to that required for global equilibrium at $T_b$.  The second stage then occurs by barrier hopping.  If, however, the system geometry (the wall positions) are set so that these ratios are equal, then the system will already have equilibrated at the end of Stage 1 and the amplitude of the Stage 2 process is zero.  This would correspond to a strong Mpemba effect.  If the difference between the two ratios is small, then one would expect the regular Mpemba effect.

\section{Are walls required?}
\label{sec:walls}

The intuitive argument linking wall anisotropy to the Mpemba effect was later made more rigorous and precise by Walker and Vucelja~\cite{Walker2022}.  One requirement for the intuitive picture to hold is a clear separation of time scales between the two stages.  This separation corresponds to $ \lambda_2 \ll \lambda_3$, with $\lambda_2 \sim \e^{-E_b / \kB T_b}$ (Kramers rate).  It is thus controlled by the barrier height, which must be large enough.  This requirement leads to a tradeoff, since data must be collected for times $\gg \lambda_2^{-1}$.  We note that the geometry ($x_\mathrm{min}$, $x_\mathrm{min}$), energy difference $\Delta E$, and barrier $E_b$ may all be set independently in an experiment.  

The above argument explicitly depends on the existence of two walls (a ``box" in the original paper).  And many further studies of the Mpemba effect for continuous systems in one dimension end up using walls, too.  For example, the two-stage relaxation scenario for the Mpemba effect~\cite{Chetrite2021} naturally leads one to wonder whether there are other mechanisms, as well.  In particular, the time-scale separation in a continuous system is most easily accomplished using a double-well potential.  Is it possible to identify a Mpemba effect using only a single well?  Several studies have shown Mpemba effects using potentials with only a single well; their common feature is the existence of two walls that are asymmetrically positioned~\cite{Walker2021,Biswas2023a,biswas2025mpemba}.

Building on these studies, recent work by Liu et al. has gone further and claimed that walls are essential for the Mpemba effect in one-dimensional potentials~\cite{Liu2026,Liu2026a}.  The argument is based on a study of the generic analytic structure of the function $a_2(T_0)$ for confining potentials $U(x)$ of polynomial form.  For such potentials, they show that the ``population transfer'' from one well to another as a function of $T_0$ for the initial states is driven by walls.  Indeed, without walls, the asymptotic behavior of $U(x) \sim x^m$ for some order $m$ is inherently symmetric and cannot lead to the asymmetric populations needed to observe the Mpemba effect.

The focus on walls is useful.  For example, it sheds light on somewhat overlooked experiments by A. Kumar in his 2021 PhD thesis that reported a Mpemba effect for a symmetric double-well potential bounded by two symmetric walls~\cite[Chapter 6]{Kumar2022a}.  As Liu et al. argue, a symmetric potential with $U'(x)=0$ at the symmetry axis is equivalent to a ``half-system'' defined for $x>0$ with an implicit wall at $x=0$.  Both systems have zero current at $x=0$~\cite{Liu2026a}.  This insight led to new results, as it was used to predict and characterize analytically a Mpemba effect in 2d radially symmetric ``ring'' potentials~\cite{Hayakawa2026}.  The radial equations again have an implicit wall at radius $r=0$.

Thus, a variety of arguments show that walls, explicit or implicit, can lead to a Mpemba effect. 

\begin{figure}
 \centering      \includegraphics[width=0.9\textwidth]{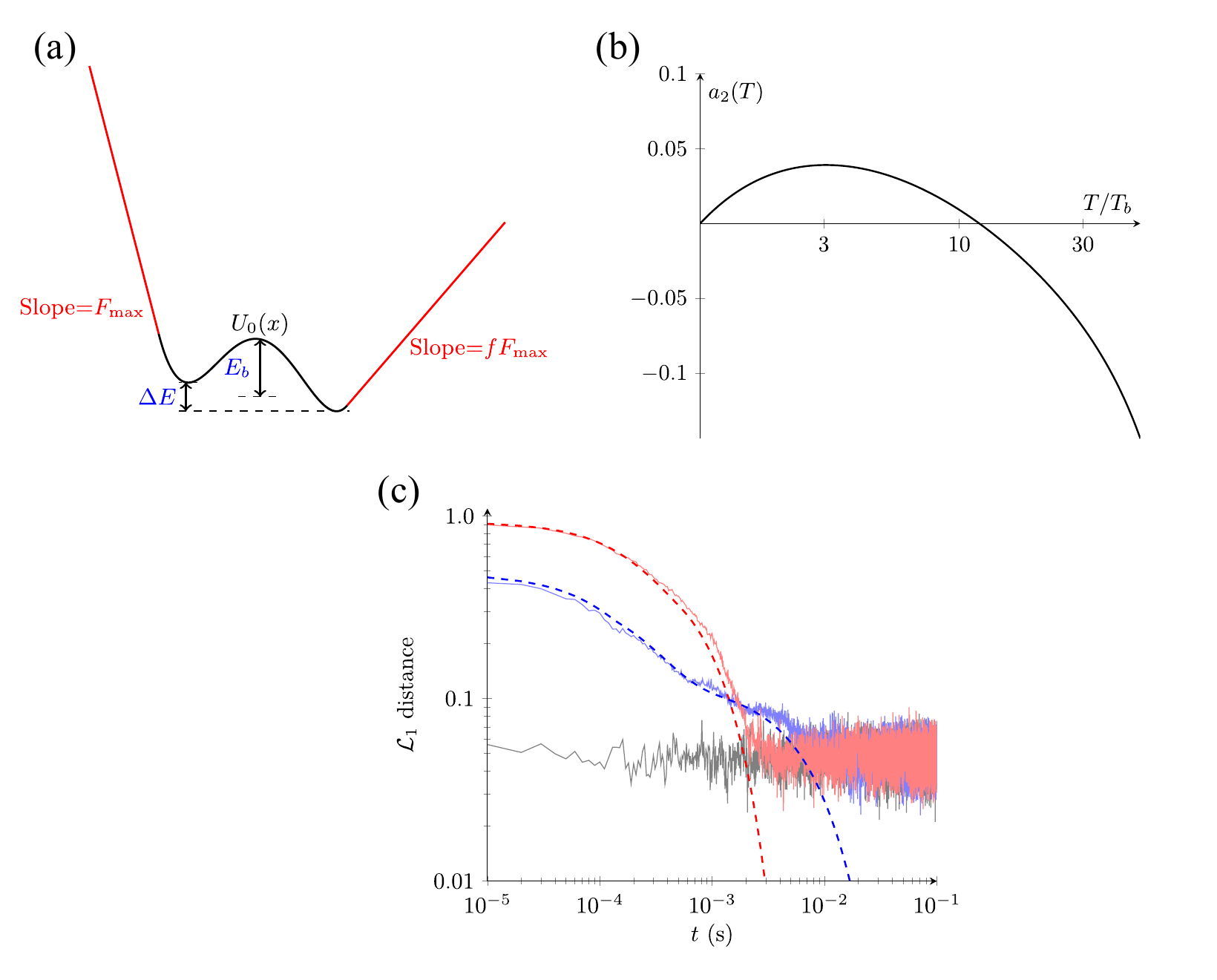}
 \caption{Mpemba effects in a potential with no box.  (a) Form of the potential, with colored segments showing the piecewise analytic sections, with the tilted double-well segment $U_0(x)$ in black and with the linear slopes in red.
 (b) The coefficient $a_2(T)$ for the potential crosses the $T$-axis at $T_0 = T_b$ and $\approx 12~T_b$.  (c) Simulations of the experiment with $N=5000$ Langevin integrations, $T_h=12~T_b$ (red), and $T_w=3~T_b$ (blue). The crossing is visible above the noise floor. Solutions to the FPE are overlaid in dashed lines. (d) Experimental data with the same parameters as the simulation. 
} 
\label{fig:asymmetricForce}
\end{figure}

\section{A potential without walls} 
\label{sec:newPotential}

In figure~\ref{fig:asymmetricForce}(a), we show a potential that is asymptotically linear at high energies.  It has different slopes but no walls, and yet, as shown in figures~\ref{fig:asymmetricForce}(b)--(d), there is a Mpemba effect.  The form of the potential is given by
\begin{align}
    U(x) &= \begin{cases}
        U_0(x), & x\in [x_l, x_r]\\
        F_{\max}(x-x_l)+U_0(x_l), & x<x_l\\
        -fF_{\max}(x-x_r)+U_0(x_r), & x>x_r
    \end{cases}\\[3pt]
    U_0(x) &= E_b\left(1- x^2\right)^2+\frac 12 \Delta E x \,,
    \label{eq:asymmetric_force_potential}
\end{align}
with $x_l$ and $x_r$ chosen as before so that $U(x)$ is $\mathcal C^1$-continuous. The $x$-coordinate is scaled by the distance between the barrier and wells for $U_0(x)$.  The form is only a slightly modification from the original shape, with the walls replaced by linear forces, with a stronger force $F_\mathrm{max}$ for $x < x_l$, the side of the higher-energy well, and a weaker force $fF_\mathrm{max}$ for $x > x_r$, the side of the lower-energy well.  The constant $f$ satisfies $0 < f < 1$ and controls the existence and type of Mpemba effect.

Like the old potential, the new one imposes a high-temperature population anisotropy between coarse-grained left and right states.  Its two asymptotic slopes (ratio $f$) in the linear regions play the same role that the wall anisotropy $x_\mathrm{min} / x_\mathrm{max}$ plays for the potential with walls.  To see this, we note that the asymptotic forms of the Boltzmann distribution are proportional to $\e^{-|x| / \ell_L}$ for $x \to -\infty$ and $\e^{-x / \ell_R}$ for $x \to +\infty$, where the decay lengths $\ell_L = \kB T_0 / F_\mathrm{max}$ and $\ell_R = \kB T_0 / fF_\mathrm{max}$.  For $T_0 \gg E_b$, distribution is approximately bi-exponential, meaning that $\Pi_L / \Pi_R \approx f$.  Thus, $f$ plays the same role as the wall anisotropy. 

\section{Requirements for experimental realization}
\label{sec:experimental}

Although the potential defined in equation~\eqref{eq:asymmetric_force_potential} does not have walls and does lead to a Mpemba effect that is easily confirmed in simulations, figure~\ref{fig:asymmetricForce}(c), it does not follow that it can be realized for the colloidal-particle study of reference~\cite{Kumar2020}.  The main issues are as follows:
\begin{enumerate}
\item The experimental distributions $p(x,t)$ must ``fit'' within experimental limits on $x$, for all $t$.
\item The crossing of distance curves computed from trajectories must be above the noise floor.
\end{enumerate}
We discuss these two constraints below.

\subsection{Domain size}
\label{sec:DomainSize}

The advantage of using a potential with infinite walls for the initial condition was that the support of the distribution at $t=0$ was confined to the interval $( -x_\mathrm{min}, x_\mathrm{max})$.  Formally, the support of potentials with ``soft walls" such as equation~\eqref{eq:asymmetric_force_potential} is infinite.  In practice, for exponential decays, it suffices for the experimental domain (about 300~nm in reference~\cite{Kumar2020}) to be several times larger than the sum of decay lengths, $\ell_L + \ell_R$.  Since the decay lengths $\sim$ initial temperature and since we are considering anomalous cooling (and not heating, which would correspond to an inverse Mpemba effect~\cite{Lu2017,Kumar2022}), it is enough to impose these constraints on the initial Boltzmann distribution, $\pi(x;T_0)$ and to consider the highest initial temperature.  A reasonable choice for the latter is to seek a strong Mpemba effect, where $a_2(T_0,T_b) = 0$.  Since our bath temperature is fixed to room temperature, this is a constraint on $T_0$.  In reference~\cite{Kumar2020}, the strong Mpemba effect occurred at $T_0 \approx 100~T_b$.  Here, the constraint on initial temperatures will require $T_0 \approx 10~T_b$.\footnote{
The numerical limit for $T_0$ depends on $F_\mathrm{max}$, which is set by the maximum laser power and the experimental limit on domain size, which is set by the properties of the position detector, a quadrant photodiode.}

\subsection{Crossing relative to noise floor}
\label{sec:crossing}

Experimentally, we detect the Mpemba effect by looking for crossing in distance curves generated by relaxations of a ``hot'' initial temperature $T_h$ and a ``warm'' initial temperature $T_w$.  As suggested, we can choose a hot temperature $a_2(T_h)=0$ to correspond to a strong Mpemba effect.  For the warm temperature, the lowest value would be defined by the start of the Mpemba effect, where the region, we recall, is defined by $\dv*{a_2}{T_0}<0$.  Thus, we can define a warm temperature $T_w$ as the lowest $T_0 > T_b$ that solves $\dv*{a_2}{T_0} = 0$.

However, we must also consider the noise floor in an experiment relative to the crossing distance.  This is set by the distance between estimates of distributions that are formed from individual time series once the system has equilibrated to the bath.  This is a function both of the form of the distribution $\pi(x;T_b)$ but, more relevant, the number of trajectories (trials) $N$ performed in the experiment and typically scales as $N^{-1/2}$~\cite{Kumar2020,Sane2026}.  In the colloidal experiments, each trial takes $\approx 0.1$ s, and ensembles of $10^3$ to $10^4$ are straightforward to obtain, in runs that last several minutes.  For larger numbers of trials, experimental drift can lead to systematic errors that can dominate the uncertainty estimates. 

Although we lack a precise criterion for predicting the crossing for particular parameter values, we observe that a larger value of $|a_2(T_w,T_b)|$ will boost the distance curve at the crossing.  Thus, we sought to maximize $|a_2|$ at $T_0=T_w$.

\subsection{Satisfying the constraints}
\label{sec:constraints}

The potential defined in equation~\eqref{eq:asymmetric_force_potential} has parameters the slope ratio $f$, which controls the ratio of well probabilities at $T_0=T_h$, the energy difference, $\Delta E$, which controls the well probability ratio at $T_b$, and the barrier height $E_b$, which controls the ratio $\lambda_3 / \lambda_2$.   The problem of choosing the parameters is complicated by cross-effects.  For example, reducing $f$ both increases the probability ratios at $t=0$, which can be useful, and increases the decay length $\ell_R$, which is not helpful.  Thus, we explored the parameter space by trial and error.  The best parameter set found so far is $f = 0.3$, $\Delta E = 0.6~\kB T_b$, and $E_b = 2~\kB T_b$, which leads to $T_h = 12~T_b$ and $T_w = 3~T_b$, which clearly displays a strong Mpemba effect with 5000 particles. These are the values used to generate figure~\ref{fig:asymmetricForce}(c).  

The typical separation $x_s \approx$ 60~nm between the two wells defines a length scale for the experiment.  In our apparatus, nearly all ($\approx 0.9996$) of the probability for the Boltzmann distribution at the highest temperature ($T_0 = 12 T_b$) is contained within the experimental domain limit ($\approx 500$~nm).

\section{
Allowable forms of potentials}
\label{sec:analytic}

As we have mentioned, our results might seem at odds with those of Liu et al.~\cite{Liu2026a,Liu2026}.  The issues are subtle, and in this section, we try to reconcile the apparently conflicting conclusions.  Perhaps most important, all agree on the main qualitative physical insight, which is the need for different asymptotic behavior for the left and right wells, in particular a steeper potential on the side of the shallow well (left in our examples).  This leads to the required variation of relative well populations as a function of initial temperature.  Liu et al. do allow for soft walls of the form $x^m$ with $m$ an even integer, but they require the slope of the asymptotic region to be steeper than the slopes in the interior double-well part of the potential and also that $m \ge 2$.  The latter condition is required for their analysis methods to be valid but leaves open cases such as ours, with $m=1$.  

\begin{figure}
    \centering
    \includegraphics[width=0.9\linewidth]{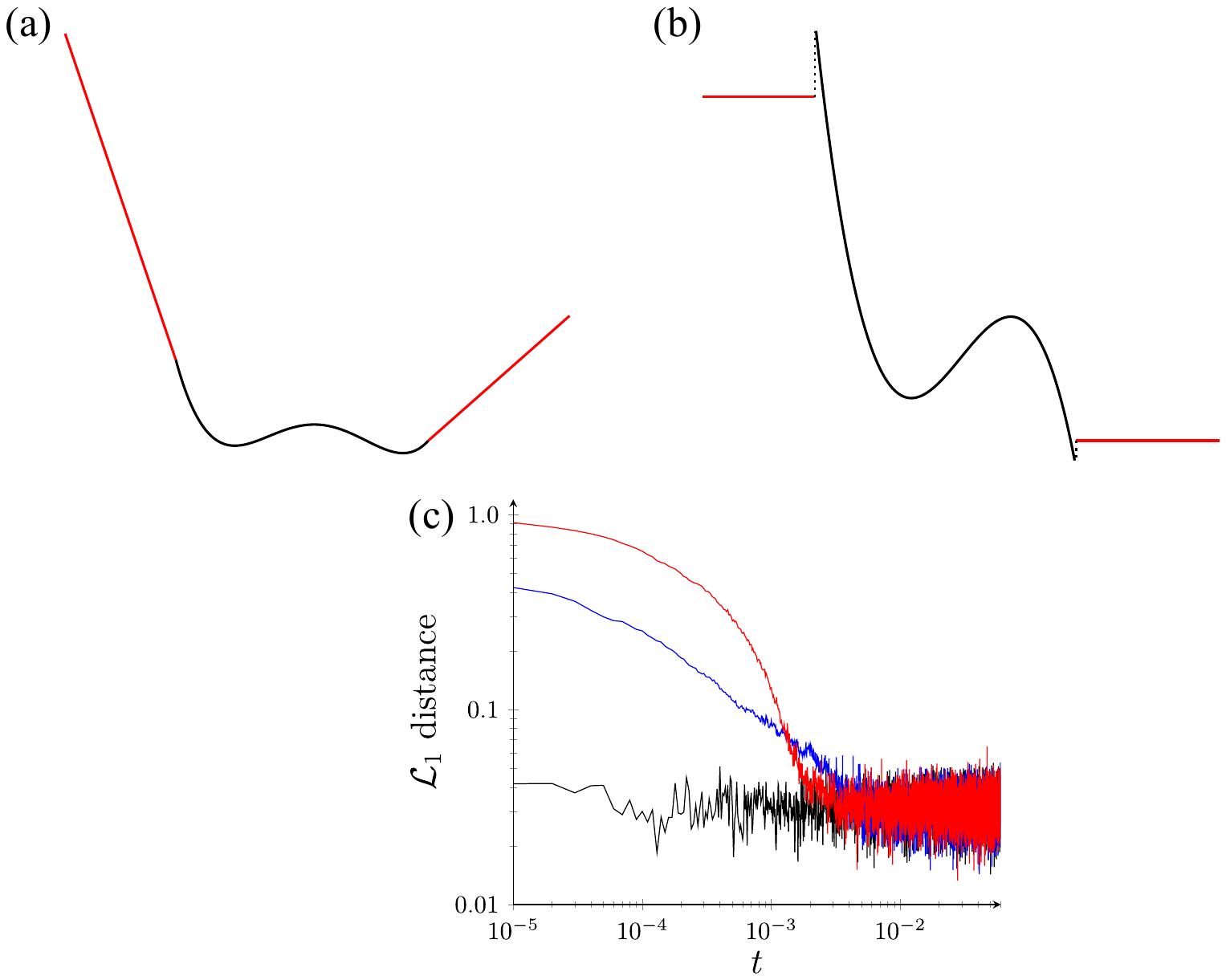}
    \caption{A bent potential can also display the Mpemba effect. (a)
    The bent potential has slopes that are discontinuous at the transition from red to black lines. (b) Force field has corresponding jump discontinuities.  (c) Crossing of distance curves indicates the Mpemba effect. 
    }
    \label{fig:bentPotential}
\end{figure}

The slopes of the outside linear regions can also be lower than values reached in the inner part of the potential. In figure~\ref{fig:bentPotential}, we reduce the asymptotic force in equation~\eqref{eq:asymmetric_force_potential} on both sides by a constant factor $g=0.8$, making the potential $\mathcal C^1$-discontinuous at the red-to-black transitions in (a).  The jump discontinuity in force (b) is more apparent. Nonetheless, simulations (c) show that the system continues to display the Mpemba effect.

Although Liu et al. allow for soft walls that have different asymptotic behavior, they view this as requiring one to open the ``Pandora's box" of non-analytic potentials.  And, indeed, potentials with walls and also potentials such as the one given here in equation~\eqref{eq:asymmetric_force_potential} have non-analytic points ($x_l$ and $x_r$ in that case).

Are non-analytic potentials non-physical?  First, we recall that many situations in classical physics assume, at least implicitly, sudden non-analytic changes in forces.  For example, rigid objects such as hard spheres have zero (or finite) interactions until they touch, where the forces suddenly increase, to large if not infinite values.  Or light refracts sharply at interfaces between different media, and so on.  But these models, while convenient, are just approximations, and their discontinuities would be smoothed out if examined at more microscopic length scales.  Still, it would be worrisome if non-analytic potentials were required to observe the Mpemba effect.  But this is not the case.

\begin{figure}
\centering    
\includegraphics[width=0.9\textwidth]{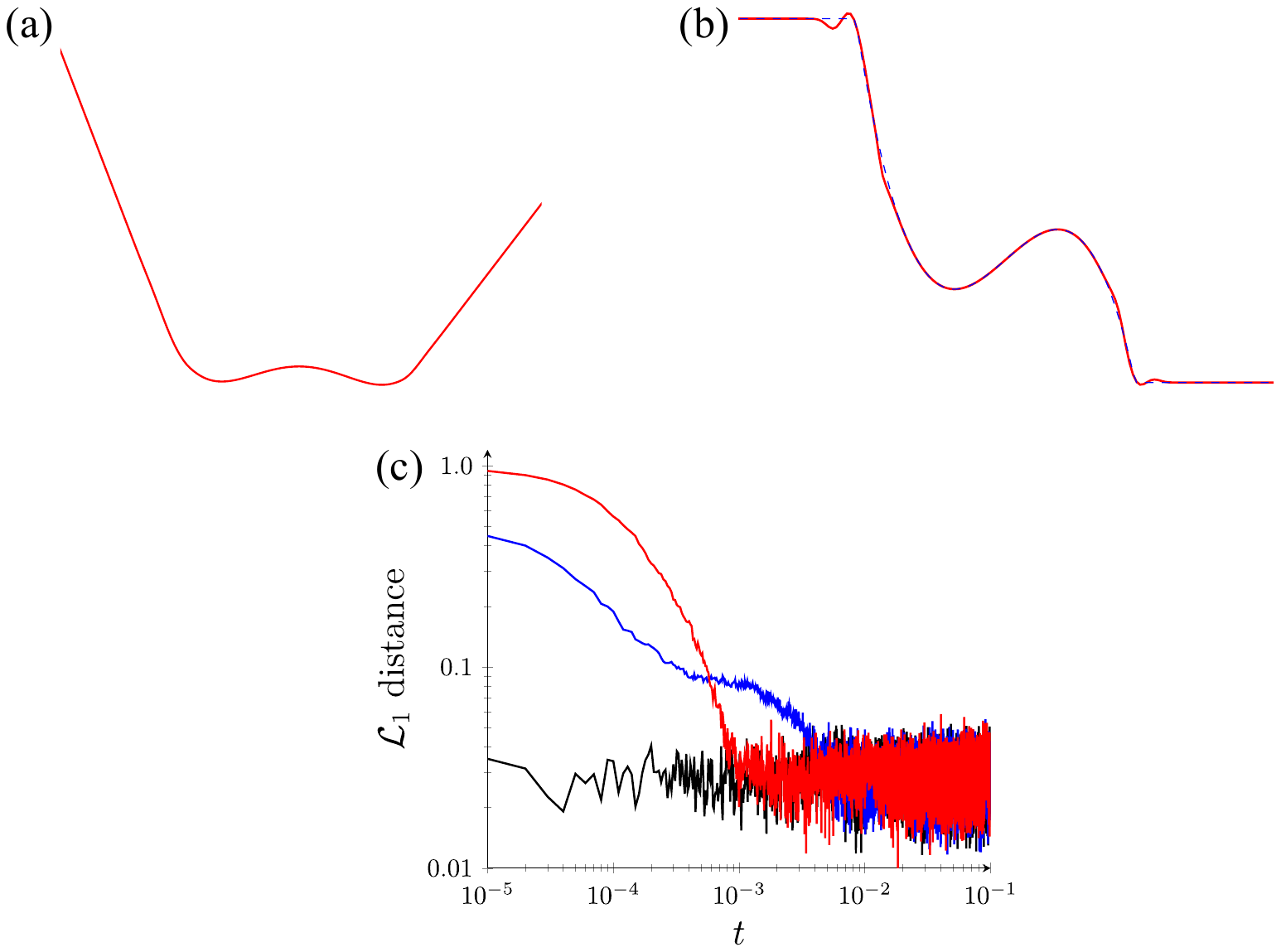}
 \caption{Fully analytic potentials on $\mathbb R$ can display the Mpemba effect. (a) Analytic approximations to the potential equation~\ref{eq:asymmetric_force_potential}, as outlined in equation~\ref{eq:analyticPotential}. (b) The force profile, $F(x)=-U_a'(x)$. In (a) and (b), the solid red lines denote the analytic potential $U_a(x)$ and the dashed blue lines the piecewise potential $U(x)$. (c) Relaxation dynamics for a Langevin simulation with the analytic potential $U_a(x)$.
} 
\label{fig:analyticPotentials}
\end{figure}

To test whether analytic potentials can show the Mpemba effect, we first rewrite equation~\ref{eq:asymmetric_force_potential} as a single-line definition:
\begin{align}
	U(x) = \mathrm{rect}\left( \frac{2x-X_r-X_l}{X_r-X_l} \right)U_0(x)
		+\mathrm{ReLU}(fF_{\max}(x-X_r))+\mathrm{ReLU}(-F_{\max}(x-X_l)) \,.
\end{align}
Here, we have used two \textit{non-analytic} functions to define $U(x)$, the \textit{rectangle} function and the \textit{Rectified Linear Unit} function:
\begin{align}
    \mathrm{rect}(u) &= \begin{cases}
        1, u\in [-1,1]\\
        0, u\notin [-1,1]
    \end{cases}\\[3pt]
    \mathrm{ReLU}(u) &= \begin{cases}
        u, u > 0\\
        0, u \leq 0
    \end{cases}.
\end{align}
The parameters $X_l$ and $X_r$ are, as before, chosen to ensure $\mathcal C^0$ continuity. To construct an analytic approximation $U_a(x)\approx U(x)$, then, we need only construct analytic approximations to the two nonanalytic functions.
\footnote{Constructing non-analytic functions or distributions as the limit of a series of analytic functions is familiar in physics: the Dirac delta $\delta(x)$, for example, can be expressed as the limit of a sequence of progressively narrower Gaussians $\lim_{\sigma\rightarrow 0}\e^{-x^2/(2\sigma^2)}/\sqrt{2\pi\sigma^2}$.} One way of approaching this is as follows:
\begin{align}
	\mathrm{rect}_n(u) &= \e^{-u^{2n}} \\
	\mathrm{ReLU}_n(u) &= \frac 1n\ln(1+\e^{nu}) \,.
\end{align}
We see that while
\begin{align}
	\lim_{n\rightarrow\infty}\mathrm{rect}_n(u) &= \mathrm{rect}(u)\\
	\lim_{n\rightarrow\infty}\mathrm{ReLU}_n(x) &= \mathrm{ReLU}(x) \,,
\end{align}
meaning that the \textit{limit} is non-analytic, even though the functions are analytic for finite $n\in\mathbb N$. Thus, we can simply replace the non-analytic functions in the potential definition with their analytic counterparts and use some sufficiently large value of $n$. We could alternatively choose a low value of $n$ but leave $X_l$ and $X_r$ as free parameters. We then sample $U(x)$ and choose $X_l$ and $X_r$ using a curve fit as parameters that make $U_a(x; X_l, X_r)$ very close to $U(x)$. We use this technique to construct the potential in figure~\ref{fig:analyticPotentials}, with
\begin{subequations}
\begin{align}
	U_a(x) &= \mathrm{rect}_n\left( \frac{2x-X_r-X_l}{X_r-X_l} \right) U_0(x)
		+ \mathrm{ReLU}_n(fF_{\max}(x-X_r)) + \mathrm{ReLU}_n(-F_{\max}(x-X_l)) \\
	U_0(x) &= E_b\left[ 1-\left(\frac x{x_w}\right)^2 \right]^2 
		+ \frac 12 \Delta E \left( \frac x{x_w} \right) \,,
\label{eq:analyticPotential}
\end{align}
\end{subequations}
with $E_b=2~\kB T_b$, $\Delta E=0.3~\kB T_b$, $f=0.5$, $F_{\max}=50$, $x_w=0.5$, and $n=5$. With these analytic approximations, we find that the analytic potential $U_a(x)$ closely matches the piecewise potential $U(x)$ from figure~\ref{fig:analyticPotentials}(a) and the $T_h$ value found by calculating that $a_2(T_h)=0$ is approximately the same.\footnote{
Simpler but somewhat ad hoc analytic potentials can also be found, such as $U(x) = \ln(1+\e^{20 \alpha (x-b)}) + \ln(1+e^{-20(1-\alpha)(x+b)}) + E_b \e^{-(x/x0)^2}((1-x^2)^2-a x)$ with $E_b = 4$, $a=0.5$, $b = 0.75$, $x_0 = 0.66$, and $\alpha = 0.33$.}

 If non-analyticity were necessary for a potential to display the Mpemba effect, we would expect that the $\mathcal L_1$ distance would never display a crossing. 
 However, we see a clear crossing in figure \ref{fig:analyticPotentials}. Thus, a potential need not be nonanalytic nor have infinite walls to display the Mpemba effect.  We use piecewise analytic potentials  then for convenience, with no measurable consequences for the relaxation dynamics.

Liu et al.~\cite{Liu2026} used a 
definition of analyticity that also 
assumed 
an infinite radius of convergence on $\mathbb C$. This definition excludes any potential that is not a pure polynomial, which is significant because $U(x) \sim x^m$ has symmetric asymptotes for $x \to \pm \infty$ and thus does not lead to a Mpemba effect for a tilted double-well potential. Liu et al.\ demonstrated that no pure polynomial potential can demonstrate the Mpemba effect; rather, the potential must either have walls, be defined piecewise, or (as we have shown) use non-polynomial functions such as logarithms, exponentials, or inverses. Crucially, the differing asymptotic behaviors required for the Mpemba effect can be generated using an analytic potential.

Thus, the different conclusions between this work and that of references~\cite{Liu2026a,Liu2026} trace back to different criteria for analyticity, having an infinite or finite radius of convergence---with polynomial or non-polynomial forms, respectively. We know of no physical reason to exclude the latter.

\section{Finite-rate quenches}
\label{sec:finiteRate}

As discussed in the Introduction, the original motivation for the work presented here was to study the Mpemba effect for finite-rate quenches imposed by a (local) bath temperature $T_b(t)$.  The theoretical motivation was to bridge between observations of Mpemba effects for instantaneous quenches from $T_0$ to $T_b$ and the fact that a Mpemba effect is not possible for quasistatic transformations between the initial and final temperatures.  The practical motivation was that extended samples have an intrinsic cooling time that depends on their geometry.  

Here in section~\ref{sec:quenchesMacroscopic}, we discuss the theoretical and practical motivations in more detail and review previous work.  Next in section~\ref{sec:exponentialQuench}, we discuss why quenches with exponential relaxation are a natural choice.  In section~\ref{sec:implementTimeDependentQuenches}, we discuss some of the experimental issues that must be addressed in order to create time-dependent effective temperatures.  Finally in section~\ref{sec:firstResults}, we present first results of numerical simulations of finite-rate quenches.

\subsection{Quenches of macroscopic systems}
\label{sec:quenchesMacroscopic}

Physically, the approach taken thus far corresponds to an \textit{instantaneous} quench, a system that is initially at equilibrium in a bath of temperature $T_h$ and starts relaxing after the bath instantaneously cools to $T_b$. However, real systems usually exist in baths that gradually cool, so that the assumption that the bath temperature drops instantly, potentially by several orders of magnitude, is in general not satisfied.  In an extended, bulk object, heat must diffuse from the bath into the bulk through a boundary, meaning that the temperature inside lags the bath temperature. 

Experiments on bulk systems that show Mpemba effects must therefore be able to cool rapidly.  For example, a recent study of Mpemba-like effects in metallic glasses used a differential scanning calorimeter to impose cooling rates of order $10^4$~K/s~\cite{Song2026}.  Ultimately, cooling rates as high as $10^{10}$~K/s have been achieved through \textit{splat cooling}, for example by using a shockwave to shoot liquid metal through a hole in a crucible, spraying little droplets onto a cold substrate~\cite{Ruhl1967}.  But a more fundamental understanding of what cooling rates are required would be helpful.

We begin by reviewing relevant work on time-varying protocols.
\begin{enumerate}
	\item Teza et al.~considered how coupling to an external bath through a ``boundary term'' affects anomalous relaxation dynamics~\cite{teza2023relaxation}.  They considered a chain of spins, with an end spin coupled to a bath and instantaneously quenched.  The other spins are coupled only energetically to the boundary (and not to the bath at all).  The gradual propagation of the boundary cooling is analogous to our single particle at finite quench rates.
	
	\item A number of studies have imposed a sequence of instantaneous quenches, with the last bath temperature being the desired ``final'' temperature of the system.  The best-known version of such a protocol is known as the \textit{Kovacs effect}, which denotes over- or undershooting of the quench temperature due to memory effects in amorphous system such as polymer melts~\cite{kovacs1964transition} or in orientational dynamics of nanoparticle suspensions~\cite{Ibanez2026}.  A particularly relevant instance of this effect was explored by Militaru et al.~\cite{Militaru2021}, who used an optically levitated nanoparticle with near-critical damping. 
	
	\item Gal and Raz considered a system that could heat more rapidly after an initial \textit{cooling} protocol~\cite{Gal2020}. They considered a four-state system which demonstrated that precooling the bath could cause the system to heat faster.

	\item The \textit{Pontus-Mpemba effect}, introduced by Nava and Egger, tries to find systems that not only cool faster but take less time to heat~\cite{Nava2025,Peluso2026}. Their protocol consists of having a system start at bath temperature $T_b$, equilibrating at temperature $T_h$, and then cooling again to $T_b$. Thus, a full thermal cycle is completed faster by a system that reaches a higher temperature.

	\item Santos has developed a minimal model for the Mpemba effect based on the Newton's-law approximation of cooling $\partial_t T = -\alpha_T (T(t-t_d)-T_b(t))$, for a system with memory~\cite{Santos2025}. Here, the ``delay time'' $t_d$ encodes the memory of past states, $T_b(t)$ is a time-dependent bath temperature, and $\alpha_T$ is the thermal diffusivity of the system. 
\end{enumerate}

With these various motivations, we consider how the dynamics of our system change if the bath temperature slowly decays with time.  A naive approach is to simply make the diffusivity $D = \kB T_b/\gamma$ a function of time:
\begin{align}
	D(t) = \frac{\kB T_b}{\gamma}h(t; T_0) \,,
\end{align}
where $h(t; T_0)>1$ is the dimensionless \textit{quench function}, the ratio of the bath temperature at time $t$ to the equilibrium temperature, $h(t; T_0) = T(t)/T_b$, given that the system started at temperature $T_0$. The quench function is always a function of the initial temperature $T_0$ at which the initial Boltzmann distribution is set to.  Here, we will focus on exponential quench functions, but one could define $h$ to also incorporate the other time-dependent protocols discussed above.

An exponential quench function also corresponds to Newtonian cooling and is given by
\begin{align}
	h_{\mathrm N}(t;T_0, \Lambda) = 
	\begin{cases}
		\frac {T_0}{T_b}, & t \leq 0 \\[5pt]
		\left( \frac{T_0}{T_b} - 1 \right)\e^{-\Lambda t} + 1, & t > 0
	\end{cases} \,.
\label{eq:newtonianQuench}
\end{align}
This quench function has a quench rate $\Lambda$ that controls the rate of cooling.

Note that the time dependence of $D(t)$ on the right-hand side of the Fokker-Planck equation 
 invalidates the form of the eigenfunction expansion in equation~\eqref{eq:FPEsoln} and thus also the Lu-Raz condition $\dv*{a_2}{T_0} < 0$ to determine the existence of a Mpemba effect given a nonzero quench rate.  We also note that as $\Lambda\rightarrow\infty$, $h_{\mathrm{N}}$ approaches the instantaneous quench,
\begin{align}
	h_{\mathrm{I}}(t;T_0) = 
	\begin{cases}
		\frac{T_0}{T_b}, & t \leq 0 \\[5pt]
		1, & t > 0
	\end{cases}.
\label{eq:instantaneousQuench}
\end{align}
Similarly, as $\Lambda\rightarrow 0$, $h_{\mathrm{N}}$ approaches a \textit{quasistatic} quench, where
\begin{align}
	h_{\mathrm{Q}}(t;T_0) = T_0/T_b \,.
\label{eq:quasistaticQuench}
\end{align}
and the system is in equilibrium at every instant in time. 

We can choose a potential that displays a Mpemba effect under an instantaneous quench ($\Lambda \gg 0$). Since no potential can display the Mpemba effect under a sufficiently slow quench ($\Lambda \rightarrow 0$), we deduce that some critical $\Lambda^*$ exists below which the Mpemba effect exists, and above which there is no Mpemba effect. 

As discussed in the introduction, a finite-rate quench is incompatible with walls:  even if we draw an initial condition contained within the walls, the subsequent evolution at high temperatures (for a long time, if the quench is slow) will lead to substantial numbers of particles leaving the domain defined by the walls and therefore invalidating the assumption of a boundary condition at the walls.  Thus, to explore finite-rate quenches, it is essential to find potentials without walls, as presented in section~\ref{sec:newPotential}.

\subsection{Motivating an exponential quench}
\label{sec:exponentialQuench}

\begin{figure}
 \centering
        \includegraphics[width=0.9\textwidth]{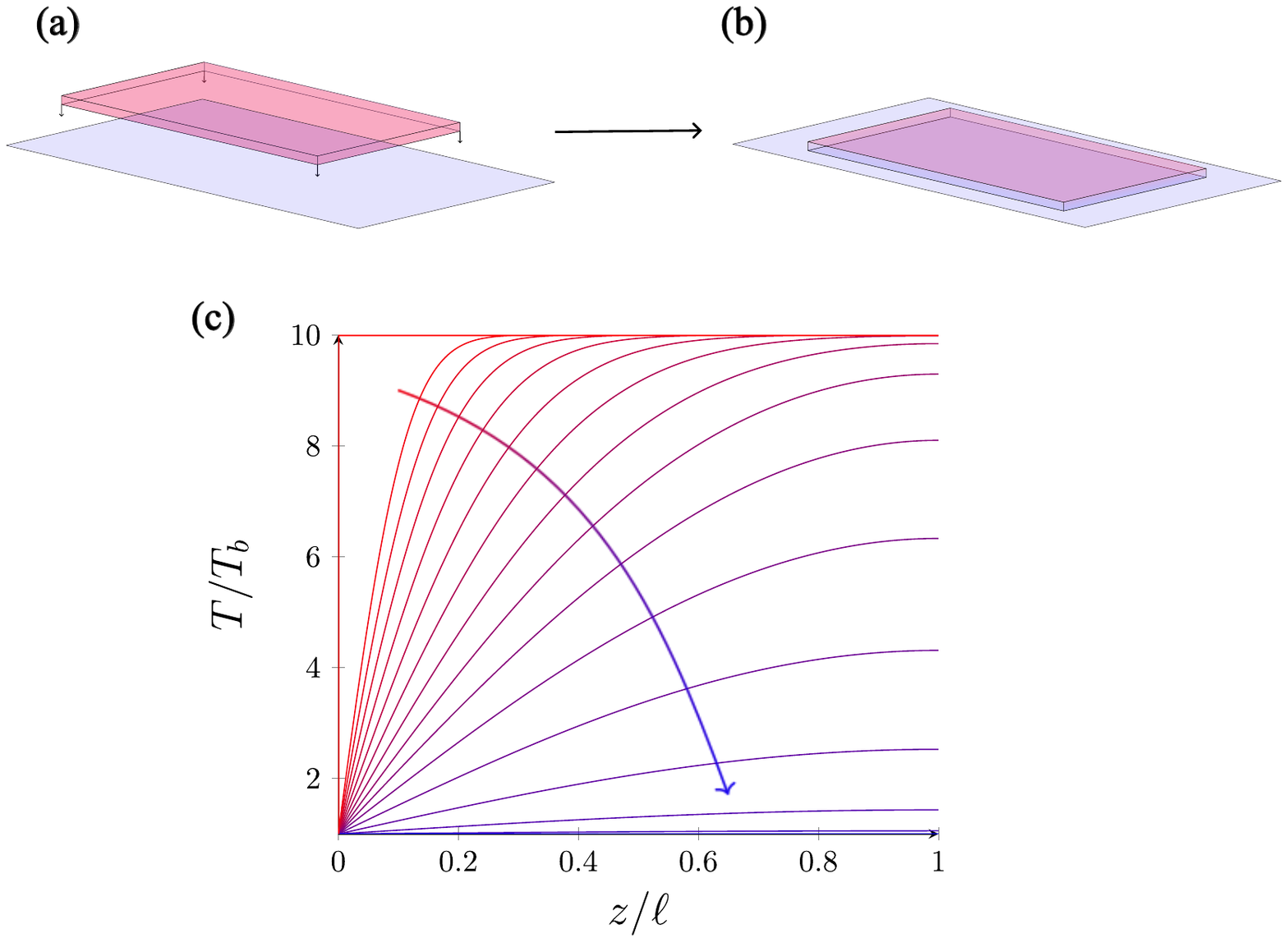}
 \caption{Model of a finite-rate quench. A one-dimensional system lives on a plate of thickness $\ell$ and uniform temperature $T_h$. (a)~At $t=0$, the plate is placed atop a much thicker base, which is a thermal reservoir held at temperature $T_b$. (b) As the plate cools to $T_b$, heat flows downwards. (c)~Temperature as a function of the normalized height $z/\ell$ at various times, for a quench from $T_h=10$ to $T_b=1$.  Redder curves represent times closer to $t=0$. 
 }
\label{fig:finite_rate_quench}
\end{figure}

Here, we argue that exponential quenches appear naturally in many experimental setups.  Consider an infinitely wide plate of thickness $\ell$ and temperature $T_0$ held above an infinitely wide base whose thickness is $\gg \ell$ and whose temperature is $T_b$, as shown in figure~\ref{fig:finite_rate_quench}(a) that is at equilibrium at temperature $T_0$.  The plate has thermal diffusivity $\alpha_T$. At $t=0$, the plate is placed atop the base, which is a heat reservoir.\footnote{
We assume the thermal diffusivity of the base $\alpha_T' \gg \alpha_T$, so that its temperature $T_b$ can be considered constant over the timescale of cooling.} 
Heat will then flow uniformly along the $z$ axis through the plate until the surface containing the system has cooled from temperature $T_0$ to temperature $T_b$. To find how the temperature of the plate evolves, we solve the heat equation with boundary conditions:
\begin{subequations}
	\begin{align}
	\pdv{T}{t} = \alpha_T\pdv[2]{T}{z} \\
    T(0, t \geq 0) &= T_b \\
	T(z > 0, 0) 			&= T_0 \\
	\eval{\pdv{T}{z}}_{z=\ell} 	&= 0.
    \end{align}
\end{subequations}
The first boundary condition enforces that the XY plane is a thermal reservoir. The second enforces the initial temperature of the system. The final boundary condition enforces that no heat flows beyond $z=\ell$:  one may imagine that the entire setup lives inside a perfectly insulating medium. 

Solving, we find that 
\begin{align}
    T(z,t) &= T_b + \frac 2\ell (T_0-T_b)\sum_{n=0}^\infty \left[ \frac{\sin\left( \frac{(2n+1)\pi z}{2\ell} \right)}{\frac{(2n+1)\pi}{2\ell}} \right]\e^{-\frac{(2n+1)^2\pi^2 \alpha_T t}{4\ell^2}}.
\end{align}
Notice, when $t=0$, the series expansion is identical to the Fourier series of a square wave of height $T_0-T_b$, which represents the initial condition that the plate is at a uniform temperature $T_0$.

Defining
\begin{align}
	\Lambda_{2n+1} = \frac{(2n+1)^2\pi^2 \alpha_T}{4\ell^2} = (2n+1)^2\Lambda_1 \,,
	\qquad  \Lambda_1 \equiv \frac{\pi^2}{4} \frac{\alpha_T}{\ell^2} \,,
\end{align}
we see that the series can be represented as
\begin{subequations}
\begin{align}
	T(z,t) &= T_b + \frac 2\ell (T_0-T_b) \left[ \frac{\sin \left( \pi z/2 \ell \right)}{\pi/2 \ell}
			\e^{-\Lambda_1 t} + \frac{\sin \left( 3\pi z/2 \ell \right)}{3 \pi/2 \ell} 
			\e^{-9\Lambda_1 t} + \dots \right] \\
   		&= T_b + 2(T_0-T_b) \left[ \frac{\sin\left( \pi z/2\ell\right)}{\pi/2}\e^{-\Lambda_1 t} 
			+ \frac{\sin \left( 3\pi z/2\ell \right)}{3 \pi/2}\e^{-9\Lambda_1 t} + \dots \right] \,.
            \end{align}
\end{subequations}
We evaluate this equation at $z=\ell$ to get the temperature of the system, $T(t)=T_bh(t)$:
\begin{subequations}
\begin{align}
	T(\ell,t) - T_b 
		&= \frac 4\pi (T_0-T_b) \left[ \sin \left( \pi /2 \right) \e^{-\Lambda_1 t} 
			+ \frac 13 \sin \left( 3 \pi/2 \right) \e^{-9\Lambda_1 t} + \dots \right] \\
		&= \frac 4\pi (T_0-T_b) \left[ \e^{-\Lambda_1 t} - \frac 13 \e^{-9\Lambda_1 t} 
			+ \frac 15 \e^{-25 \Lambda_1 t} - \dots \right] \,.
            \end{align}
\label{eq:heatEqn}
\end{subequations}
The term in the brackets has the form of a series $x-x^9/3+x^{25}/5-\dots$, which is identical to the Taylor series expansion of $\arctan(1)=\pi/4$ when $t=0$, thus canceling the prefactor. Since $\Lambda_3 = 9\Lambda_1$, the second term is expected to be much smaller than the first for sufficiently large times, and the series may be truncated to one term at sufficiently large times, yielding 
\begin{align}
	T(\ell,t) -T_b \sim (T_0-T_b) \, \e^{-\Lambda_1 t} \,.
\end{align}
Dividing throughout by $T_b$, we find the expression in equation~\eqref{eq:newtonianQuench},
\begin{align}
	h(t; T_0, \Lambda_1)-1 \sim (T_0/T_b -1) \, \e^{-\Lambda_1 t} \,.
\end{align}
Since $\Lambda_1 \sim \alpha_T / \ell^2$ depends only on the thermal diffusivity $\alpha_T$ and plate thickness $\ell$, we may vary it independently of the system eigenvalues $\lambda_n$, thus quenching the system at a speed limited only by its size.

Thus, we see how in a simple, representative geometry, time-dependent quenches naturally tend to an exponential form, at least at time scales comparable to the overall quench rate.  Although our argument was for a one-dimensional geometry, it leads to a result that matches a naive dimensional argument ($ \ell^2/\alpha_T$ has units of time).  One might then expect to find similar exponential quenches in other geometries, where $\ell$ is the smallest geometric scale of the system.

\subsection{Experimental implementation of time-dependent quenches}
\label{sec:implementTimeDependentQuenches}

As we have discussed, we cannot use the initial-value method to implement finite-rate quenches; however, there are several methods for adding external random forces that lead to effective temperatures that produce static thermal distributions that are identical to those of high-temperature systems.  

An early implementation of this technique to colloidal systems applied a random voltage to electrodes.  The voltage values applied to the electrodes were random numbers drawn from a Gaussian distribution with mean zero and a variance set by the desired effective temperature.  The resulting time-dependent electrical fields in the fluid then lead to random forces on particles, as long as the particle or the fluid has electrical charges~\cite{Martinez2013}.   A later variation replaced the computer-generated random numbers with analog random voltages generated by the Johnson noise of a resistor at room temperature that is amplified by an analog electrical circuit~\cite{Saha2023}.  Yet another variation by Chupeau et al. uses random displacements (produced by an acousto-optical deflector) of the optical tweezer that traps particles~\cite{Chupeau2018}.  As long as displacements are small enough to stay within the linear range of the trap, a Gaussian random number generator can produce Gaussian-distributed random forces.  Using any of these methods should allow for effective temperatures $T_0 \lesssim 10 T_b$.

\subsection{First results on simulations of finite-rate Mpemba transitions}
\label{sec:firstResults}

\begin{figure}
 \centering
        \includegraphics[width=0.9\textwidth]{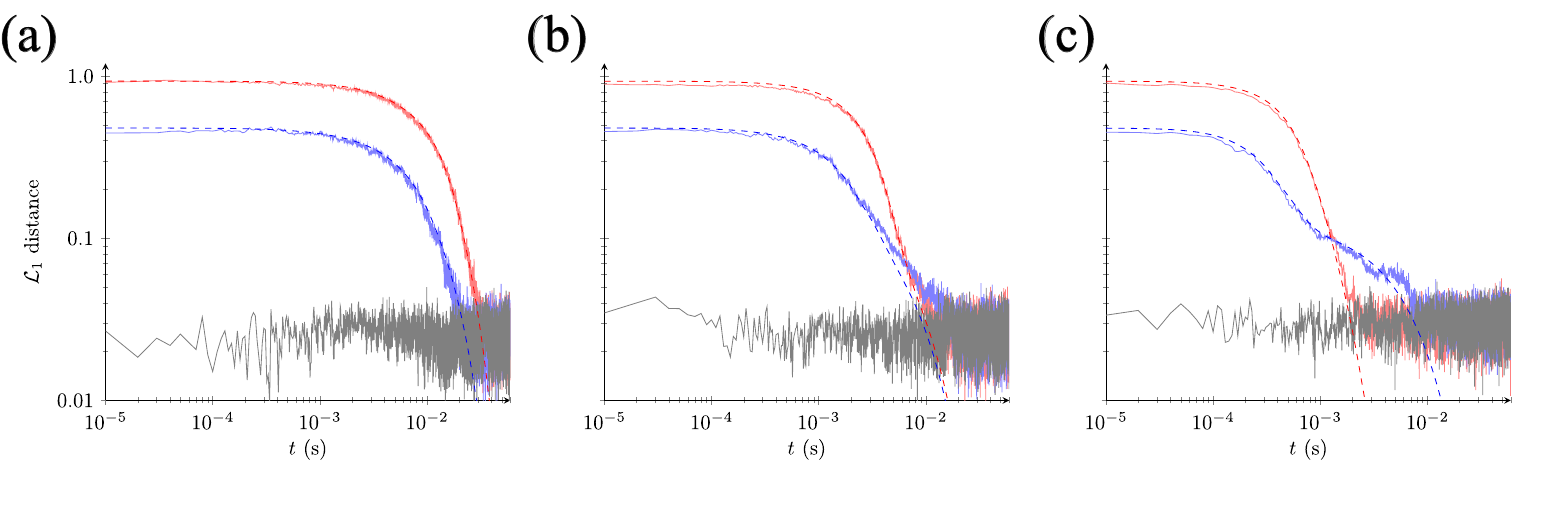}
 \caption{Numerical simulations show a critical quench rate for a Mpemba effect to exist. (a) Fast quench rate, $\Lambda = \lambda_3$. (b) Approximate critical quench rate, $\Lambda = 0.27~\lambda_3 \approx \Lambda^*$. (c) Slow quench rate, $\Lambda = \lambda_2$. Solutions to the FPE are overlaid using dashed lines.
 }
\label{fig:finiteQuenchSim}
\end{figure}

In Langevin simulations 
of finite-rate transitions for a regular Mpemba effect, we used an exponential quench of the form of equation~\eqref{eq:newtonianQuench} with quench rates $\Lambda$ that were fast ($\Lambda = \lambda_3$), slow ($\Lambda = \lambda_2$) and intermediate ($\lambda_2 < \Lambda < \lambda_3$).  

Figure~\ref{fig:finiteQuenchSim} shows distance-decay curves for slow, intermediate, and fast quench rates at temperatures for the potential equation~\eqref{eq:asymmetric_force_potential} with $T_h=12~T_b$ and $T_w=3~T_b$ for $N = 10000$ particles. The slow quench rate, $\Lambda=\lambda_2$ clearly shows no crossing and hence no Mpemba effect, consistent with expectations for the quasistatic limit. The fast quench rate $\Lambda=\lambda_4$ clearly shows a crossing that indicates a Mpemba effect, consistent with the expected result for infinite quench rate as shown in figure~\ref{fig:asymmetricForce}(c). (The quench rate $\Lambda = \lambda_2 \approx 0.058~\lambda_3$ is larger than the quasistatic limit, which would require $\Lambda \ll \lambda_2$.)  Finally, the intermediate rate $\Lambda \approx 0.27~\lambda_3$, is close to the critical rate $\Lambda^*$ to just have a (detectable) Mpemba effect.

\section{Conclusions}
\label{sec:conclusions}

To summarize, we have
\begin{itemize}

\item resolved the ``two potential'' experimental issue in work by Kumar et al.~\cite{Kumar2020};

\item showed that a box is not required to produce the Mpemba effect;

\item showed that walls and non-analytic potentials do lead to Mpemba effects and are convenient but that the Mpemba effect can be produced by potentials that are analytic and have finite radius of convergence in the complex plane;

\item showed first results on finite-rate transitions that demonstrate a critical quench rate to achieve a Mpemba effect.

\end{itemize}

The immediate goal of future work is to test the scenarios outlined here experimentally.  Preliminary work using the ``shaking tweezer'' method has led to significant issues with clipping of the noise when the combined potential and thermal forces becomes too large.  The alternate strategy of imposing electrical forces via external electrodes~\cite{Martinez2013,Saha2023} may be better.
By using two separate sources of forces (electrodes for the stochastic force and optical-trap displacements for the deterministic forces that create the virtual potential), one avoids the clipping problem.  As an added advantage, one can in practice create greater forces using electrodes than optical traps.  For example, Saha et al. produced electrical forces as high as $100 T_b$.  However, these authors also observed effects due to well-known electrohydrodynamic forces that are produced at low frequencies.  One can minimize these nonlinear forces by high-pass filtering the noise applied to the electrodes.  But then one is no longer applying white noise, and this change might itself lead to other sources of systematic deviations from ideal behavior.  Finding an effective strategy for producing large, accurate effective temperatures remains an open experimental issue.

On the theoretical side, our identification of a critical quench rate was visual and qualitative.  It would be better to formulate a more precise definition.  As a step in that direction, we note that for times greater than $\Lambda^{-1}$, the bath temperature approaches $T_b$, a constant.  In this state, we can define the $a_2$ coefficient unambiguously, which will be constant in this regime.  However its amplitude will be more complicated than the simple projection of the initial state with temperature $T_0$ onto the bath equilibrium solution.  Instead, the amplitude should depend on $T_0$, the form of the quench (equation~\ref{eq:newtonianQuench}), and the quench rate.  Fixing the form of the quench, we would expect there to be a function $a_2(T_0,T_b,\Lambda)$.  The critical quench rate $\Lambda^*$ would then be fixed by defining it to be the lowest value of $\Lambda$ for which there exists a temperature $T_0$ where $\pdv*{a_2}{T_0} = 0$, at fixed $\Lambda$.  One could investigate this function numerically or via analytic techniques, perhaps using tools developed for non-autonomous open quantum systems~\cite{Peluso2026} or the Magnus expansion from time-dependent perturbation theory~\cite{Blanes2010}.

Finally, we note that the ``naive'' approach to incorporating a time-dependent quench solely through a time-dependent diffusivity $D(t)$ may hide issues that should be addressed in future work.  In particular, Brey and Casado have argued that the Langevin dynamical equations will acquire extra terms beyond ones arising from explicit time dependence of the temperature~\cite{Brey1990}, as will the corresponding Fokker-Planck equation~\cite{Romero1994}.  The complications are expected to be present in the overdamped limit, too~\cite{Gomez2021}.  Whether these terms apply to our situation is unclear, as the effective temperature is created by a fluctuating external force and not by a physical change to the bath temperature.

\section*{Acknowledgments}
We acknowledge useful discussions with Hisao Hayakawa, Fr\'{e}d\'{e}ric van Wijland, and Rapha\"{e}l Ch\'{e}trite concerning their recent work \cite{Liu2026,Liu2026a,Hayakawa2026}.  We thank in particular F. van Wijland for pointing out the relevance of Ref.~\cite{Brey1990}.  This work was supported by a Discovery Grant awarded by the Natural Science and Engineering Research Council of Canada (NSERC).

\printbibliography
\end{document}